\documentclass[a4paper]{spie}  
\usepackage[]{graphicx}

\usepackage{color}

\title{Spectral calibration and modeling of \\the NuSTAR CdZnTe pixel detectors}

\author{Takao Kitaguchi\supit{a},
 Brian W. Grefenstette\supit{a},
 Fiona A. Harrison\supit{a},
 Hiromasa Miyasaka\supit{a},\\
 Varun B. Bhalerao\supit{a},
 Walter R. Cook III\supit{a},
 Peter H. Mao\supit{a},
 Vikram R. Rana\supit{a},\\
 Steven E. Boggs\supit{b},
 Andreas C. Zoglauer\supit{b}
 \skiplinehalf
 \supit{a}Caltech Division of Physics, Mathematics and Astronomy, Pasadena, USA; \\
 \supit{b}U.C. Berkeley Space Sciences Laboratory, Berkeley, CA
}

\authorinfo{Further author information: (Send correspondence to T.K.) T.K.: E-mail: kitaguti@srl.caltech.edu}

\begin{document}
  \maketitle

\begin{abstract}
 The Nuclear Spectroscopic Telescope Array (NuSTAR) will be the first space mission to
 focus in the hard X-ray ($5-80$~keV) band.  The NuSTAR instrument carries two co-aligned
 grazing incidence hard X-ray telescopes.   Each NuSTAR focal plane
 consists of four 2~mm CdZnTe hybrid pixel detectors, each with an
 active collecting area of 2~cm $\times$ 2~cm.
 Each hybrid  consists of a $32\times32$ array of 605~$\mu$m pixels,
 read out with  the Caltech custom low-noise NuCIT ASIC.
 In order to characterize
 the spectral response of each pixel to the degree required to meet
 the science calibration requirements, we have developed a model based
 on Geant4 together with the Shockley-Ramo theorem customized to the
 NuSTAR hybrid design. This model combines a Monte Carlo of the X-ray
 interactions with subsequent charge transport within the detector.
 The combination of this model and calibration data taken using
 radioactive sources of $^{57}$Co, $^{155}$Eu and $^{241}$Am enables us
to determine electron and hole mobility-lifetime products for each pixel,
and to compare actual to ideal performance expected for defect-free material.
\end{abstract}



\section{Introduction}
\label{sec:intro}  

The Nuclear Spectroscopic Telescope Array (NuSTAR)
mission\cite{2010SPIE.7732E..21H} is a NASA Small Explorer that will carry the
first focusing hard X-ray ($5-80$~keV)
telescopes to orbit.
It is scheduled to be launched into low Earth orbit (550~km altitude,
6~deg inclination) by a Pegasus XL rocket from Kwajalein Atoll
in February 2012.
NuSTAR has two co-aligned telescopes, each consisting of a depth-graded
multilayer optic\cite{2009SPIE.7437E..10K,2010SPIE.7732E..22H}
focusing onto a cadmium zinc telluride (CdZnTe) pixel detector\cite{2009SPIE.7435E...2R}.
The detector is placed at the bottom of a cylindrical active CsI anti-coincidence shield to
reduce detector background.

Cadmium zinc telluride (CdZnTe) and cadmium telluride (CdTe) semiconductor materials are attractive for use
in astronomical hard X-ray instruments because, compared to alkali-halide scintillators,
they can achieve superior spectral resolution and imaging detectors can be
implemented in compact geometries.
These materials also have relatively high atomic number (48 for Cd and 52 for Te) and a
wide enough band gap ($\sim 1.5$~eV) to permit near room temperature operation.
Because of these advantages, cosmic hard X-ray detectors using CdZnTe
and CdTe are already operating in space onboard the
Swift\cite{2004ApJ...611.1005G} and INTEGRAL\cite{2003A&A...411L.131U}
satellites, respectively. Both missions contain coded aperture instruments with a large number of
single planar detectors (32,768 CdZnTe for Swift, 16,384 CdTe for INTEGRAL).  Due to the
planar architecture, and the resulting relatively high capacitance and leakage current, and charge
collection non-uniformity, the spectral resolution of these coded mask imagers is limited.
NuSTAR will be the first astronomical mission to utilize CdZnTe hybrid pixel detectors, where
the detector anode is divided into small segments, and low electronic noise, small leakage and
good spectral resolution ($<$1~keV FWHM from $5-80$~keV) is achievable.

Exploiting the good energy resolution for spectroscopy of astrophysical sources requires
constructing a highly-accurate response model.     For both CdZnTe and CdTe, the low
hole mobility leads to a non-gaussian energy response, in particular spectral 'tailing'.
As a result, determining the resolution kernel requires a detailed
charge transport model, combined with Monte Carlo simulations.
In addition, in the case of CdZnTe,  current
high-pressure Bridgman material growth processes produce crystals
with a nonuniform distribution of charge transport properties.
This results in variation of the charge transport
properties across a centimeter-scale detector. We must therefore measure these properties for each pixel
in order to construct a NuSTAR detector response model that can meet requirements for spectral reconstruction.

In this paper, we describe the NuSTAR spectral calibration program,
the charge transport model, and we present results obtained by fitting
data from NuSTAR flight detectors.

\section{Cadmium-Zinc-Telluride Pixel Detector for NuSTAR}
\label{sec:CdZnTe}

Each NuSTAR focal plane consists of four CdZnTe hybrid pixel detectors
arranged in a $2 \times 2$ array (see Figure~4 in
[\citenum{2010SPIE.7732E..21H}]).
The sensor element in each hybrid is a CdZnTe crystal with a segmented anode.   Each anode
pixel is attached to the input pad of a custom readout circuit using a gold-wire/conductive
epoxy interconnect.   The low-noise ASIC (NuCIT \cite{1998SPIE.3445..347C}),
was developed by Caltech originally for the {\em HEFT} balloon
program\cite{2005ExA....20..131H}, and upgraded for homeland security and again for {\em NuSTAR}.
The CdZnTe crystals  were manufactured by Endicott Interconnects (formerly eV Products).
Each crystal has an active collecting area of 2 cm $\times$ 2 cm and a thickness of 2 mm.
The cathode electrode is a monolithic platinum contact and the anode
electrode is patterned into a $32 \times 32$ grid with each pixel having
pitch of 605~$\mu$m, forming 1024 pixels for each hybrid.   The pixel contacts
are separated by a  50~$\mu$m gap.
The NuCIT ASIC contains a common on-chip analog-to-digital converter and
an external microprocessor supports the readout.
Each of the 1024 (a 32 x 32 array matched in pitch to the sensor) readout circuits has its own preamplifier, shaping amplifier,
discriminator, sample and hold circuitry and test pulsar.

The NuCIT ASIC preamplifier and readout are designed so that both an electron signal
and a hole signal can be determined for each event, allowing a depth of interaction measurement.
To obtain pulse height information, the output of each pixel's preamplifier is presented as a current signal sequentially
and continuously to a bank of 16
sampling capacitors, with a $\sim 100$~ns (programmable) dwell time on each capacitor. The signal is
also presented as a voltage to a discriminator.  When a trigger occurs, sampling is halted.   At
this point the capacitor bank contains approximately eight samples ``pre-trigger'' and eight
``post-trigger''.     The integrated charge signal is obtained by subtracting the pre-trigger samples
from the post-trigger samples.   In normal operation, the instrument control processor searches all triggered pixels
for the highest pulse height and reads nine sets of sixteen pulse height samples from a
$3 \times 3$ array.  The processor packs the charge signal, obtained by performing the subtraction
described above, into an array containing the pixel with the
highest signal at the center, and the 8 surrounding pixels.   In subsequent ground processing, based on a threshold
determined in software, the surrounding pixels are sorted into electron (positive amplitude)
and hole (negative amplitude) signals, and each group is summed.  The negative amplitude
hole signal results from the trapped holes creating image charge on the anode pixels (the
holes are trapped for a timescale longer than the readout time).  In order
to cancel common mode noise, the average of the hole signal is subtracted from
the electron signal.

\begin{figure}[ht]
 \begin{minipage}{0.475\hsize}
  \begin{center}
   \begin{tabular}{c}
    \includegraphics[width=8cm]{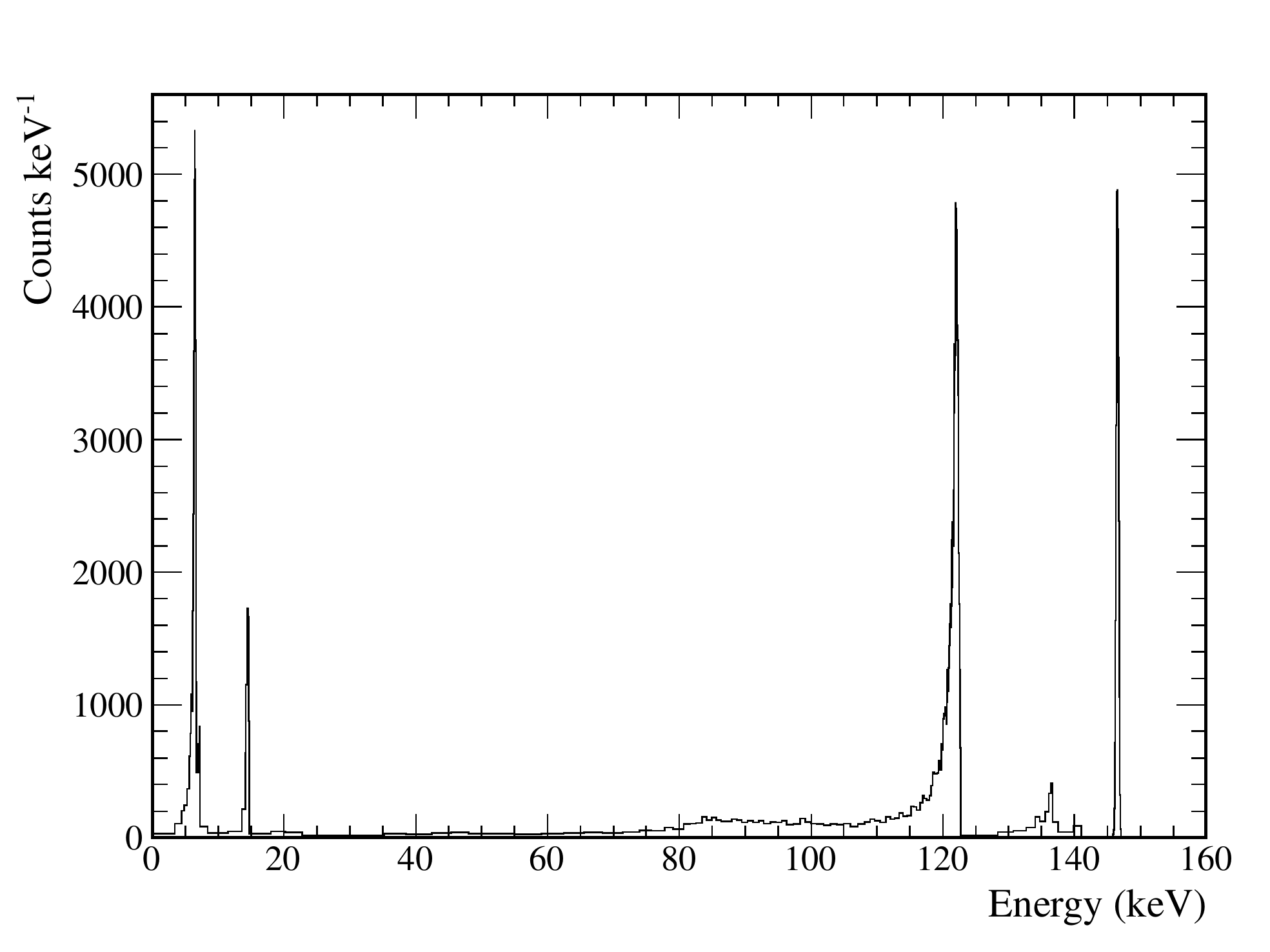}
   \end{tabular}
  \end{center}
  \caption{\label{fig:SpecCo57}
  A typical spectrum of one CdZnTe pixel obtained with a $^{57}$Co
  radioactive source, which emits strong $\gamma$-ray lines of 6.40,
  7.06, 14.4, 122 and 136 keV. The 146 keV peak is produced by
  the on-chip test pulsar.
  Events that only the center pixel is triggered are collected.
  Data were taken at 278~K temperature and -450~V high voltage for one day.}
 \end{minipage}
 \hspace{0.5cm}
 \begin{minipage}{0.475\hsize}
  \vspace{-1.3cm}
  \begin{center}
   \begin{tabular}{c}
    \includegraphics[width=8cm]{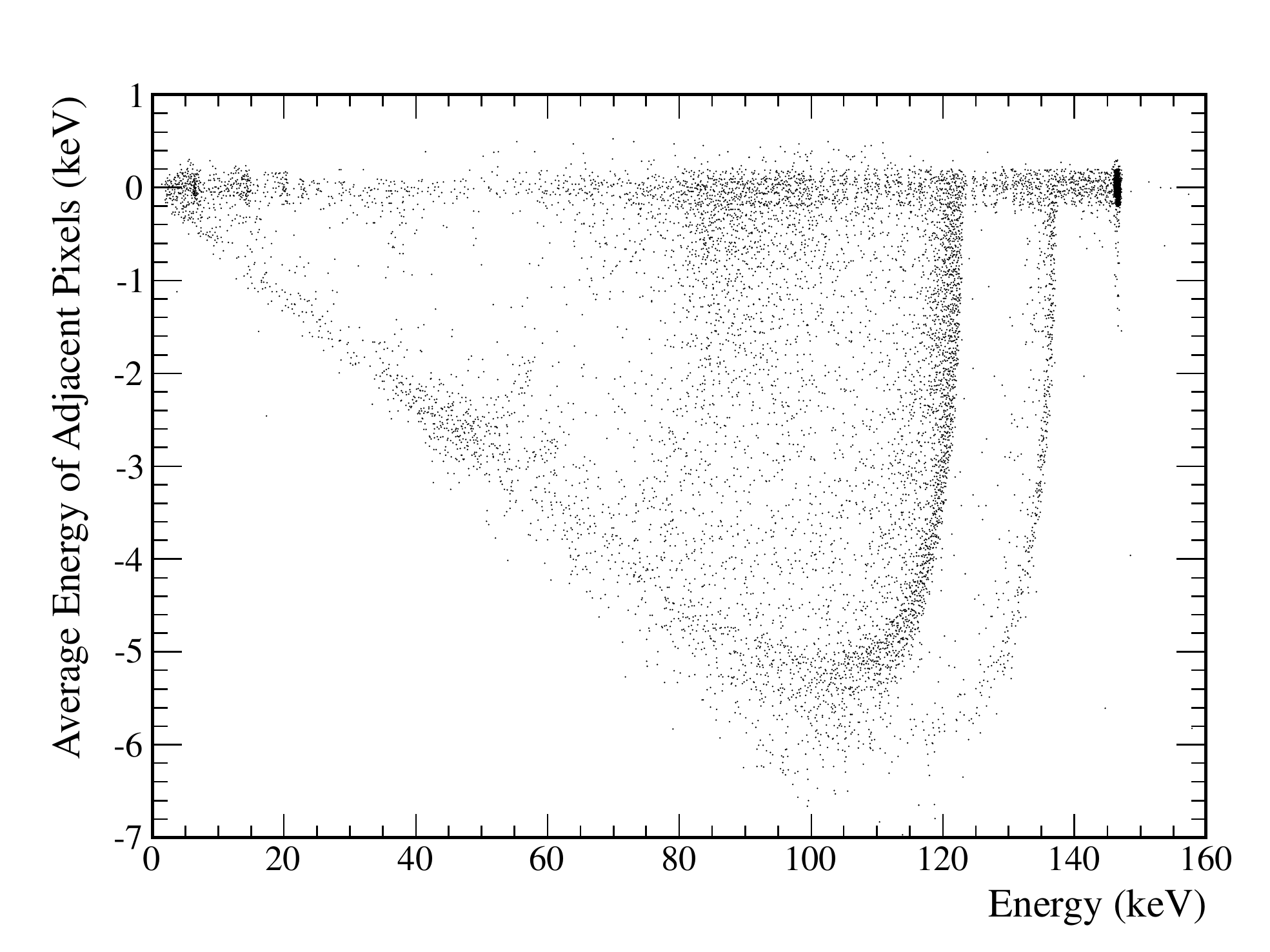}
   \end{tabular}
  \end{center}
  \caption{\label{fig:DPlotCo57}
  A typical scatter plot of the center pixel energy (electron signal)
  vs the average energy of 8 adjacent pixels (hole signal).
  Figure~\ref{fig:SpecCo57} is obtained by the projection of this
  scatter plot along the y-axis.}
 \end{minipage}
\end{figure}

Figure~\ref{fig:SpecCo57} shows a typical CdZnTe spectrum obtained with a $^{57}$Co radioactive source.
The spectrum contains events where only
the central pixel is triggered ({\em ie.} excluding events where charge is shared
among multiple pixels). The energy resolution at 14.4 and 122~keV is
0.5~and 0.9~keV FWHM respectively.
This resolution is much better than that achieved with a monolithic CdZnTe
detector (7~keV at 122~keV for Swift\cite{2005NIMPA.541..372S},
9~keV at 100~keV for INTEGRAL\cite{2003A&A...411L.131U}) due to
the small pixel effect\cite{1995PhRvL..75..156B} and the ability to remove pixel-to-pixel
charge collection variation. The test pulse, which appears at about 147~keV in the spectrum,
has a width of 0.4~keV FWHM, showing the NuCIT ASIC and leakage current noise is very low.
A low-energy tail associated with the 122~keV main peak is also seen in Figure~\ref{fig:SpecCo57}.
The tail structure is produced by a combination of hole trapping and Compton down-scattering of
the line  $\gamma$-rays in the radioactive source holder and passive material surrounding
the detector.

Figure~\ref{fig:DPlotCo57} shows a scatter plot of the center
pixel energy (electron signal) vs the average energy of the 8 surrounding pixels (hole image signal)
for single pixel events.
The events produced by 122 keV gamma-rays are largely distributed in a line tilted at an angle
relative to the y-axis.  At large values of hole signal, the event track
curves back towards the origin.   The location of an event on this track depends on
interaction depth and the charge transport properties of the pixel.
The central pixel pulse height decreases with interaction depth as a result of
the variation in the magnitude of the image charge signal
with depth.   The closer to the anode the event occurs, the greater the
image charge, and this image charge is both detected on the surrounding pixels, and
subtracts from the central pulse height.
Low-energy photons ($E < 60$~keV) interact only in a shallow region near
the cathode surface and are observed to produce very little hole signal.
However, the 122 keV photons in Figure~\ref{fig:SpecCo57} occur at a range of depths in the CdZnTe,
and so produce the ``depth curve''.
When collapsed onto the X-axis to produce a standard spectrum, this is
observed as a low-energy tail to the line.

An advantage of the depth measurement is that the CdZnTe activation background induced by cosmic radiation
can be suppressed by flagging and rejecting events with large interaction
depths.
As noted above, photons with energy $< 60$~keV that enter through the aperture interact on average near the
cathode surface, while the activation background occurs uniformly in the
CdZnTe crystal.
We can then flag and reject events that occur deep in the detector as
background events.

\section{Monte-Carlo Simulator}
\label{sec:sim}

\subsection{Overview of Simulator}
\label{sec:sim:overview}

\begin{figure}[ht]
 \begin{minipage}{0.5\hsize}
  \begin{center}
   \begin{tabular}{c}
    \includegraphics[width=8.5cm]{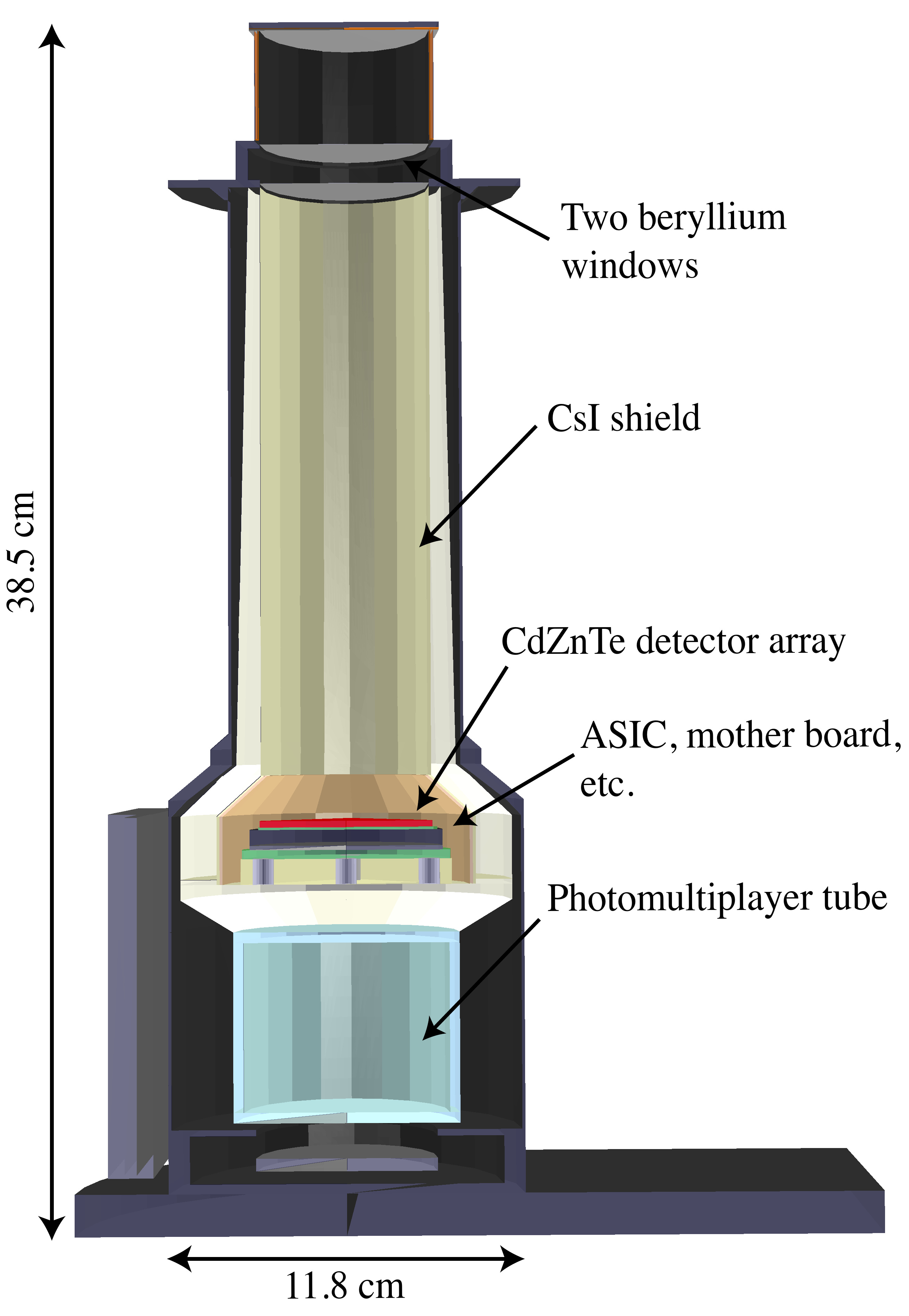}
   \end{tabular}
  \end{center}
  \caption{\label{fig:NuGeom}
  Cross-section view of the NuSTAR focal plane detector displayed by Geomega
  (ROOT-based OpenGL viewer).}
 \end{minipage}
 \hspace{0.3cm}
 \begin{minipage}{0.475\hsize}
  \begin{center}
   \begin{tabular}{c}
    \includegraphics[width=7.8cm]{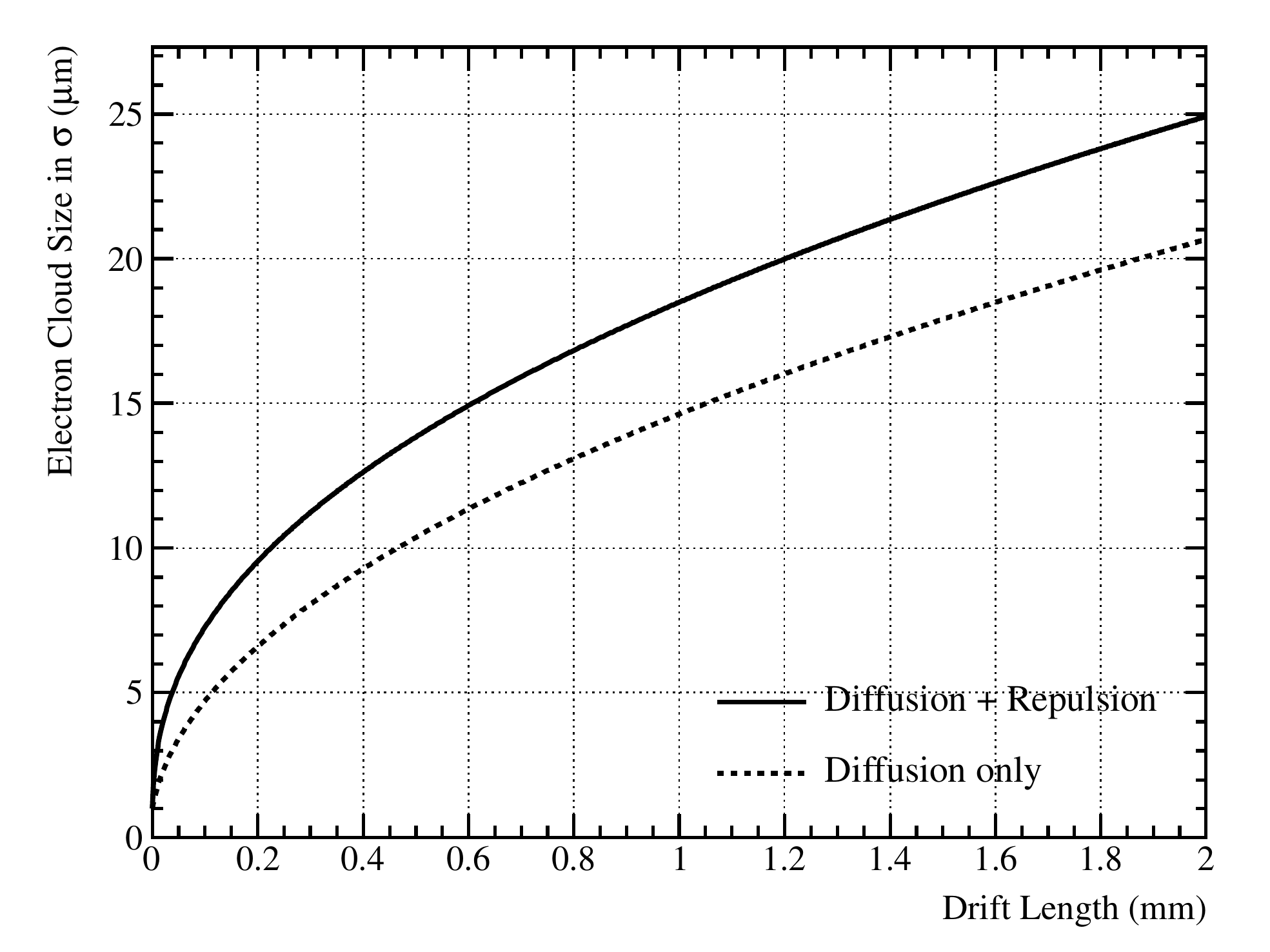}
   \end{tabular}
  \end{center}
  \caption{\label{fig:eDiff}
  Standard deviation of electron cloud size at a 60 keV energy deposition
  with an initial radius = 1~$\mu$m.
  The solid and dotted lines represent electron cloud size with and without
  electrostatic repulsion, respectively.
  The applied high voltage is 450~V and temperature is 5 degree.
  Although the electron mobility is set to be
  $10^{-3}$~cm$^{2}$~s$^{-1}$~V$^{-1}$,
  the curves are nearly independent of it.}
 \end{minipage}
\end{figure}

It is almost impossible to analytically calculate the detector response
of CdZnTe pixel detectors because the geometry of the focal plane
detector is complex, and moreover the physical processes
governing particle interactions have many channels and branches.
Monte-Carlo simulations provide the only method of calculating the
spectral response function.
In order to generate the pixel-by-pixel response,
we have developed a Monte-Carlo simulator based on the Geant4
toolkit\cite{2003NIMPA.506..250G} together with a charge transport model
customized to the NuSTAR hybrid design described in \S\ref{sec:CdZnTe}.
The NuSTAR focal plane geometry in Figure~\ref{fig:NuGeom} has been
described with Geomega, one of the
MEGAlib\cite{2006NewAR..50..629Z,2008SPIE.7011E.101Z} software
libraries designed for medium-energy $\gamma$-ray astronomy.

The simulator is operated on an object-oriented software framework
called ANL\cite{2006ITNS...53.1310O},  originally developed
for the ASCA satellite in the 1990s, and  improved for the
follow-on Suzaku and ASTRO-H missions.
The ANL framework is based on the concept that a complicated
process can be decomposed into many simple modules, and can be described
as a chain of them. Each ANL module includes an initial function,
a function for event-by-event analysis, and a termination function.
The ANL framework calls these modules in the order defined by the chain.
The Geant4 run manager and three user mandatory classes (the primary
particle generator, detector geometry constructor, and physics process
manager) are wrapped into the corresponding ANL modules separately.
Examples of the successful application of this combination of Geant4 and ANL framework
can be seen in the generation of the detector response for the hard X-ray
detector onboard Suzaku\cite{2005ITNS...52..902T} and the background
estimation for the soft $\gamma$-ray detector onboard
ASTRO-H\cite{2010SPIE.7732E.105M}.

\subsection{Charge Transport Model}
\label{sec:sim:CHmodel}

The Geant4 Monte Carlo engine generates hit information such as times,
positions, and deposited energies for each physical
interaction. After collecting the hit information, the downstream
ANL modules calculate the output charge induced on
each anode pixel, and convert them into digitized values by considering
the response of the sensor and the properties of the electronics.
The electron and hole charge transport is the most important
among the processes because the low-energy tails are
generated by charge (in particular hole) trapping.  We have therefore paid special attention to
developing an accurate and complete charge transport model.

In the first stage of the model, we calculate the fraction of charge shared
among anode pixels using the deposited energy and interaction
position.  We assume that the electron cloud can be described by a 2-d Gaussian shape on the
plane parallel to the electrode, with a standard deviation, $\sigma(t)$,
which increases as a function of time.
The electron cloud spreads due to two physical processes; random
thermal motion of electrons, and electron-electron repulsion.
According to Benoit and Hamel (2009) [\citenum{2009NIMPA.606..508B}], the diffusion equation
describing charge cloud spreading due to thermal motion and repulsive force is given by
\begin{equation}
 \label{equ:diff}
 \frac{{\rm d}\sigma(t)^{2}}{{\rm d}t}
  = \frac {2 \mu_{\rm e}k_{\rm B}T} {e}
  + \frac {\mu_{\rm e} e N(E_{\rm dep})} {12\pi^{3/2}\epsilon_{0}\epsilon_{\rm r}\sigma(t)},
\end{equation}
where $\mu_{\rm e}$ is the electron mobility,
$k_{\rm B}$ is the Boltzmann constant,
$T$ is the operating temperature,
$e$ is the elementary charge,
$N(E_{\rm dep})$ is the number of electron-hole pairs as a function of
the deposited energy,
$\epsilon_{0}$ and $\epsilon_{r}$ are the vacuum permittivity and
the relative permittivity of CdZnTe (typically 10.9), respectively.
The first and second terms on the right hand side represent diffusion
effects due to thermal motion and repulsion force, respectively.
The standard deviation at the moment electrons arrive at the anode
pixel can be obtained by solving equation~(\ref{equ:diff}) using
the Runge-Kutta method,
assuming that the initial value of the standard deviation is set to
1~$\mu$m for every energy deposition.
As show in in Figure~\ref{fig:eDiff}, the electron cloud size
is quite dependent on the repulsive force --  without including electrostatic repulsion the electron
cloud size is underestimated by $\sim20$\%.
The fraction of charge collected on each pixel is determined by integrating
the normalized 2-d Gaussian area with the obtained standard deviation
in the corresponding anode pixel region.
This charge sharing model predicts about 2/3 of events
have only one triggered pixel (single-pixel event)
and the rest have multiple triggered pixels.
The ratio is in good agreement with observed results.

Next we determine the charge induction efficiency of each pixel.
The Shockley-Ramo theorem (eg. [\citen{2001NIMPA.463..250H}]) provides
a convenient way to calculate induced charge due to the motion of
charge carriers.
The charge induction efficiency at any time, $\eta(t)$, is given by
\begin{equation}
 \label{equ:CIE}
  \eta(t)
  = \int^{ \overrightarrow{x_{\rm e}}(t) }_{\overrightarrow{x_{0}}}
  \exp \left(-
	\frac{|\overrightarrow{x}-\overrightarrow{x_{0}}|}
	     {\mu\tau_{\rm e} E(\overrightarrow{x})}
       \right)
  \nabla\phi_{\rm w} \cdot {\rm d} \overrightarrow{x}
  + \int^{\overrightarrow{x_{\rm h}}(t)}_{\overrightarrow{x_{0}}}
  \exp \left(-
	\frac{|\overrightarrow{x}-\overrightarrow{x_{0}}|}
	     {\mu \tau_{\rm h} E(\overrightarrow{x})}
       \right)
  \nabla\phi_{\rm w} \cdot {\rm d} \overrightarrow{x},
\end{equation}
where $\overrightarrow{x(t)}$ is the position at time, $t$,
$\overrightarrow{x_{0}}$ is the interaction position where electron-hole
pairs are generated,
$\mu\tau$ is the charge carrier mobility-lifetime product,
$E$ is the applied electric field,
and $\phi_{\rm w}$ and $\nabla\phi_{\rm w}$ are the weighting potential
and weighting field.
The e and h subscripts represent electrons and holes.
The first and second terms on the right hand side represent the charge
induction efficiency for electrons and holes, respectively.
The exponential terms describes carrier trapping effects.
This treatment does not consider detrapping, since that occurs on
a timescale longer than the electronics response time.
In the case of an ideal single CdZnTe detector,
which has a uniform applied electric field and a uniform weighting
field throughout the detector volume,
equation~(\ref{equ:CIE}) becomes the well-known Hecht
relation\cite{1932ZPhy...77..235H}.

We require a fine-mesh to describe the weighting potential in order to
numerically integrate equation~(\ref{equ:CIE}).
Because some electrons drift toward the adjacent pixels due to
diffusion, the weighting potential requires a
$3 \times 3$ pixel array consisting of the center and 8 adjacent pixels.
We divide the detector volume containing $3 \times 3$ pixels into 31 million
rectangular cells consisting of $360 \times 360$ parts on a side
and 240 parts in depth.
The weighting potential at each cell center is calculated by using
the approximate solution\cite{2004JVST...22..975K} assuming the cathode
electrode is an infinite plane.  We assume an anode pitch of 605~$\mu$m and a detector
thickness of 2~mm provide the boundary values for the calculation.
Each pixel is assumed to have an identical weighting
potential even if it is near the crystal edge.

We require calculation of the carrier track in order to calculate equation~(\ref{equ:CIE}).
To be exact, each charge carrier drifts along a different path
due to random thermal motion.
However, it would require long computation times to individually integrate
equation~(\ref{equ:CIE}) for every single path.
In order to shorten the computing time, we assume that all holes take
the shortest path from interaction point to the cathode electrode.
These holes are drifted in the vertical (cathode) direction
with a constant velocity given by the hole mobility multiplied
by the applied electric field,
presumed to be constant throughout the CdZnTe volume.
Electrons also take the shortest path from the interaction point to
individual anode pixels,  which have a significant amount of shared charge.
In this work, this amount is set to be more than 0.1\%, determined by our
charge sharing model.
The charge induction efficiency the anode of each pixel is
estimated by numerically integrating equation~(\ref{equ:CIE}),
using the shortest charge carrier path.

We obtain the charge sharing ratio and the charge
induction efficiency of individual anode pixels for each physical
interaction as described above.
In the final stage, we calculate the energy deposition in each pixel by multiplying
the deposited energy by the ratio and efficiency.
We then calculate the total energy of each pixel by summing up all of
the energies produced by every physical interactions in the CdZnTe detector.
Because the observed energy is broadened due to Poisson fluctuations in
the number of hole-electron pairs, and the variation in paths traveled by the charge
carriers, the final energy deposition is randomly smoothed by a Gaussian.
Using these energies, we can reproduce the  CdZnTe spectrum in Figure~\ref{fig:SpecCo57}
and the scatter plot in Figure~\ref{fig:DPlotCo57}.

\section{Measurement of Charge Transport Properties}
\label{sec:mes}

To reproduce the spectral response of each pixel,
the NuSTAR Monte Carlo simulator needs a determination of the electron and hole mobility-lifetime
products, assigned by equation~(\ref{equ:CIE}), for each
pixel.  We determine these charge transport properties by using calibration data taken with radioactive sources
with a range of line energies spanning the NuSTAR band.   The calibration data are
fit using comparison data produced by the NuSTAR Monte Carlo simulator as described below.

\begin{figure}[ht]
 \begin{minipage}{0.475\hsize}
  \begin{center}
   \begin{tabular}{c}
    \includegraphics[width=8cm]{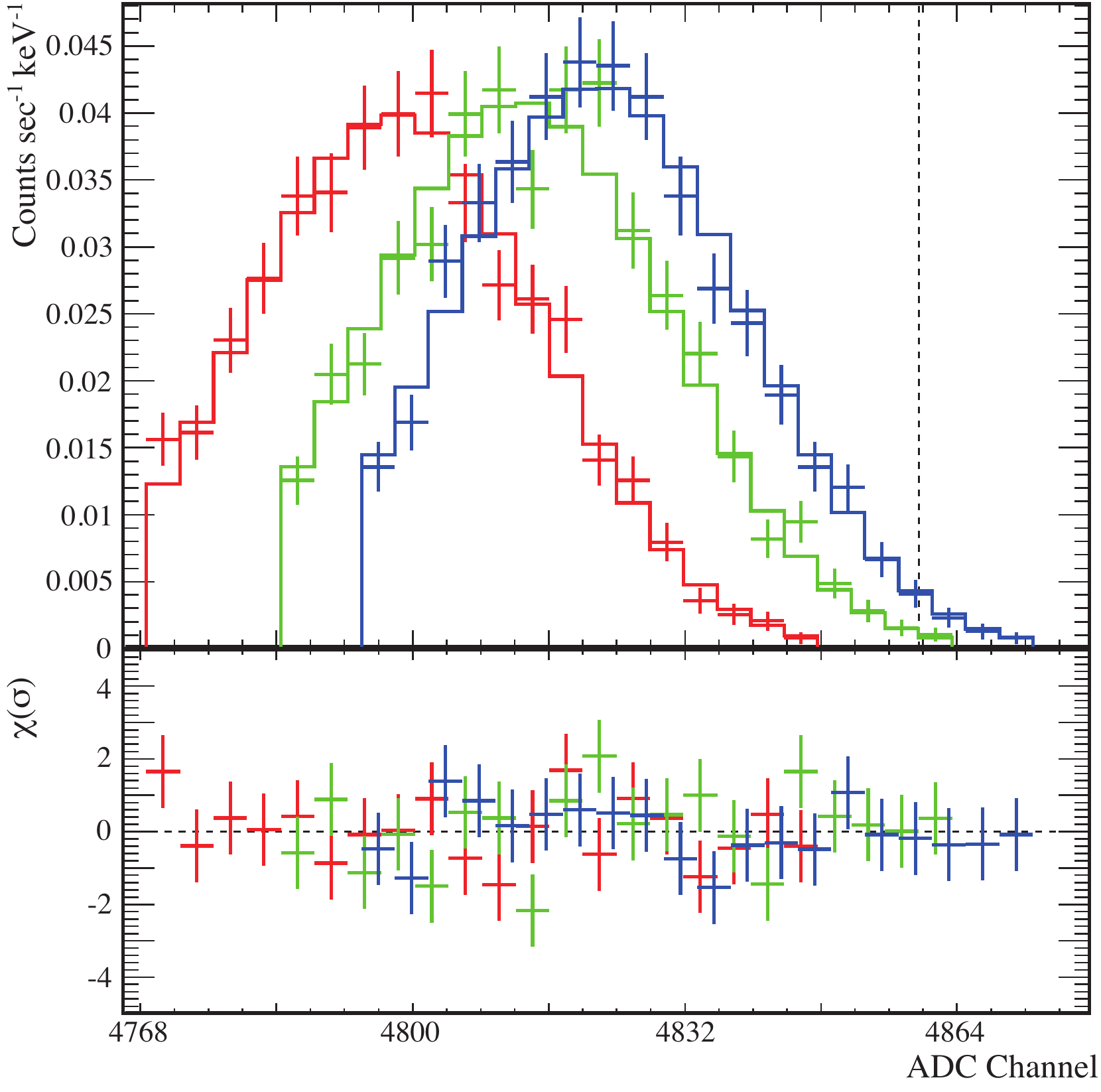}
   \end{tabular}
  \end{center}
  \caption{\label{fig:SpecHVScan}
  59.5 keV line spectra taken with $^{241}$Am at different 3 high
  voltages of 300 (red), 400 (green), and 500 V (blue).
  The vertical dotted line represents the expected peak with no
  electron trapping [Ch$_{0}$ in equation~(\ref{equ:Epeak})].
  The lower panel shows the fit residuals in term of $\sigma$.}
 \end{minipage}
 \hspace{0.5cm}
 \begin{minipage}{0.475\hsize}
  \begin{center}
   \begin{tabular}{c}
    \includegraphics[width=7.5cm]{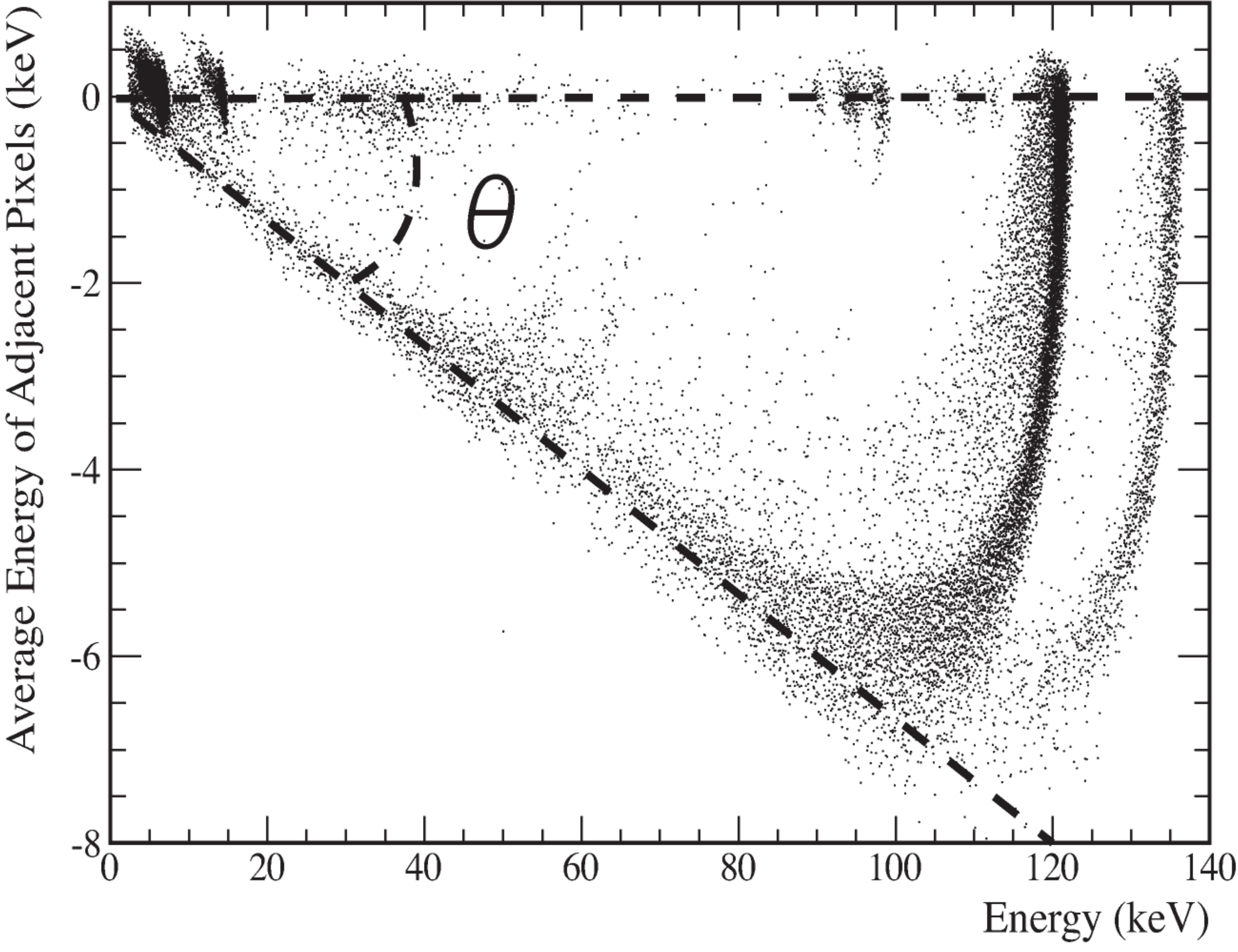} \\
    \includegraphics[width=7.5cm]{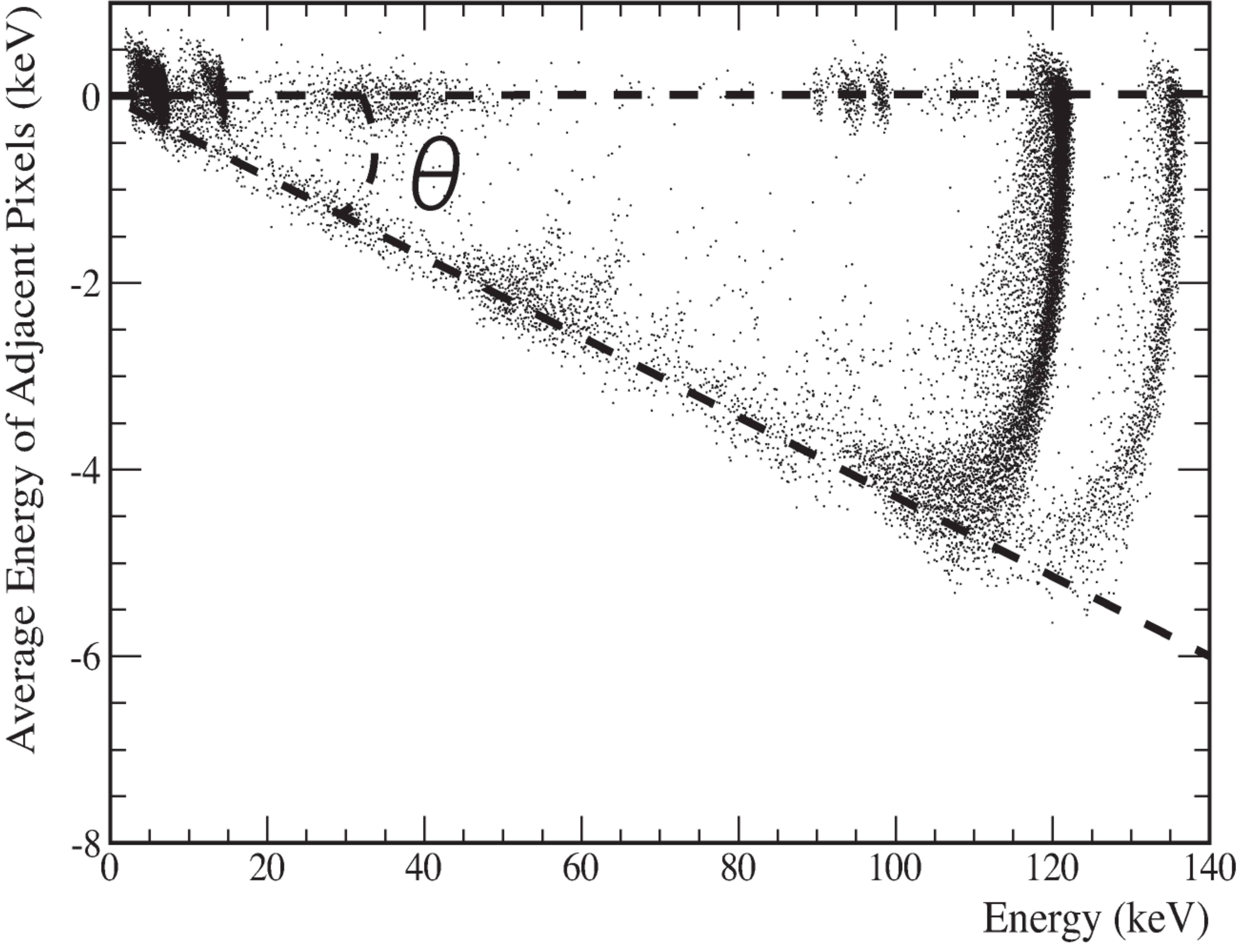}
   \end{tabular}
  \end{center}
  \caption{\label{fig:DPlots}
  Simulated depth plot irradiated with $^{57}$Co.
  $\mu\tau_{h}$ is set to $10^{-5}$ (top) and
  $3 \times 10^{-5}$ (bottom) cm$^{2}$~V$^{-1}$.}
 \end{minipage}
\end{figure}

In order to determine the mobility-lifetime product for electrons,
we take data to determine the voltage dependence of a peak channel (total charge signal),
${\rm Ch}_{\rm peak}$, given approximately by
\begin{equation}
 \label{equ:Epeak}
 {\rm Ch}_{\rm peak} \approx
 {\rm Ch}_{\rm 0} \times \exp \left( \frac{d^{2}}{\mu\tau_{\rm e} V} \right),
\end{equation}
where ${\rm Ch}_{0}$ is the peak channel with no electron trapping,
$d$ is the detector thickness (2 mm for the NuSTAR CdZnTe detector),
and $V$ is the applied high voltage.
The electron trapping, described by the exponential term,
becomes strong as the voltage decreases.
Figure~\ref{fig:SpecHVScan} shows an example of voltage scan
spectra taking with $^{241}$Am at three different voltages compared
to spectral models produced by the Monte-Carlo simulator.
The spectral peak shifts to lower energy as a function of decreasing  high voltage as expected,
with the magnitude of the shift relatively well described by the model  ($\chi^2/{\rm dof} = 40.4/54$).
The derived $\mu\tau_{\rm e}$ is
$(8.78 \pm 0.28) \times 10^{-3}$~cm$^{2}$~V$^{-1}$.
This method for measuring $\mu\tau_{\rm e}$ is the same as that for
the Swift CdZnTe detectors \cite{2005NIMPA.541..372S}, but the
weighting potential is customized for to the NuSTAR hybrid design.

In order to establish a procedure for determining the hole mobility-lifetime
product using the simulation code, we fit the energy vs. depth of interaction plot.
Figure~\ref{fig:DPlots} shows energy vs. depth scatter plots for two different
hole mobility-lifetime products.
We found that $\gamma$-rays that interact subject to a particular mobility-lifetime
produce a distribution with respect to the anode electrode with a definite angle, $\theta$, shown in
Figure~\ref{fig:DPlots}.
The absolute angle monotonically decreases as the
hole mobility-lifetime products increases.
We can therefore measure the mobility-lifetime products for holes
by comparing the observed angle to a data base of angles database calculated
by the simulator.

For determining the electron mobility-lifetime product we used bias
voltages of  300, 400 and  500~V and for each illuminated the detectors
with a $^{241}$Am source placed about 3~inch above the center of
the motherboard where the CdZnTe hybrids are installed.
We operated the detectors  at 5 $^\circ$C at atmospheric
pressure.  Each data set at was taken at a constant voltage for 12 hours.
For the hole mobility-lifetime products we obtained data with
$^{57}$Co and $^{155}$Eu radioactive sources, illuminating the detectors for 24 hours for each source.
We confirmed by simulating  data  that these measurement time periods are long enough to determine the
mobility-lifetime products for electrons and holes within the required
accuracy.

\begin{figure}[ht]
 \begin{minipage}{0.45\hsize}
  \begin{center}
   \begin{tabular}{c}
    \includegraphics[height=7.5cm]{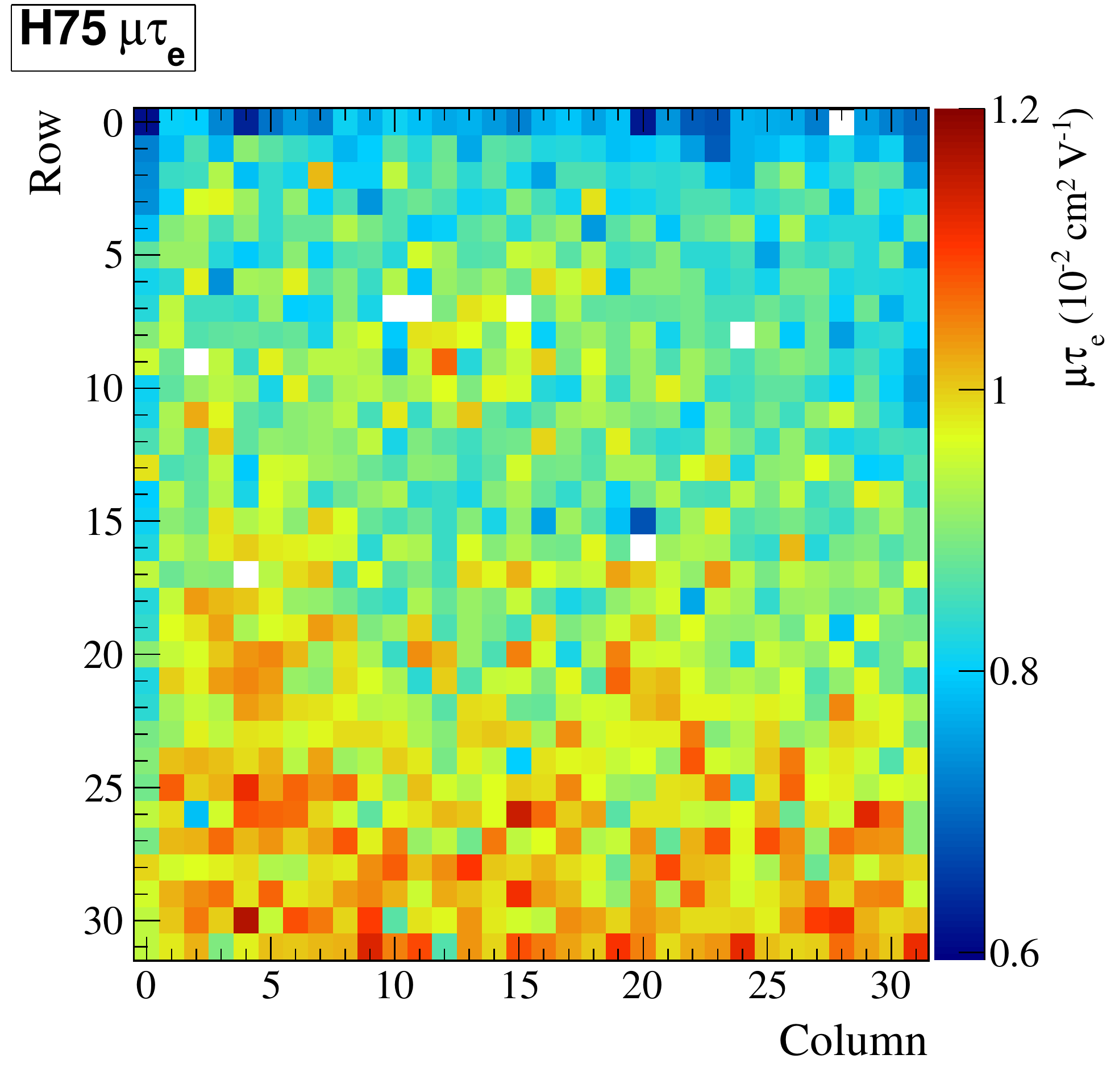}
   \end{tabular}
  \end{center}
  \caption{\label{fig:MapMuTauE}
  Map showing obtained electron mobility-lifetime products for
  the $32 \times 32$ pixel array in a CdZnTe hybrid.}
 \end{minipage}
 \hspace{0.5cm}
 \begin{minipage}{0.45\hsize}
  \vspace{-0.5cm}
  \begin{center}
   \begin{tabular}{c}
    \includegraphics[height=7.5cm]{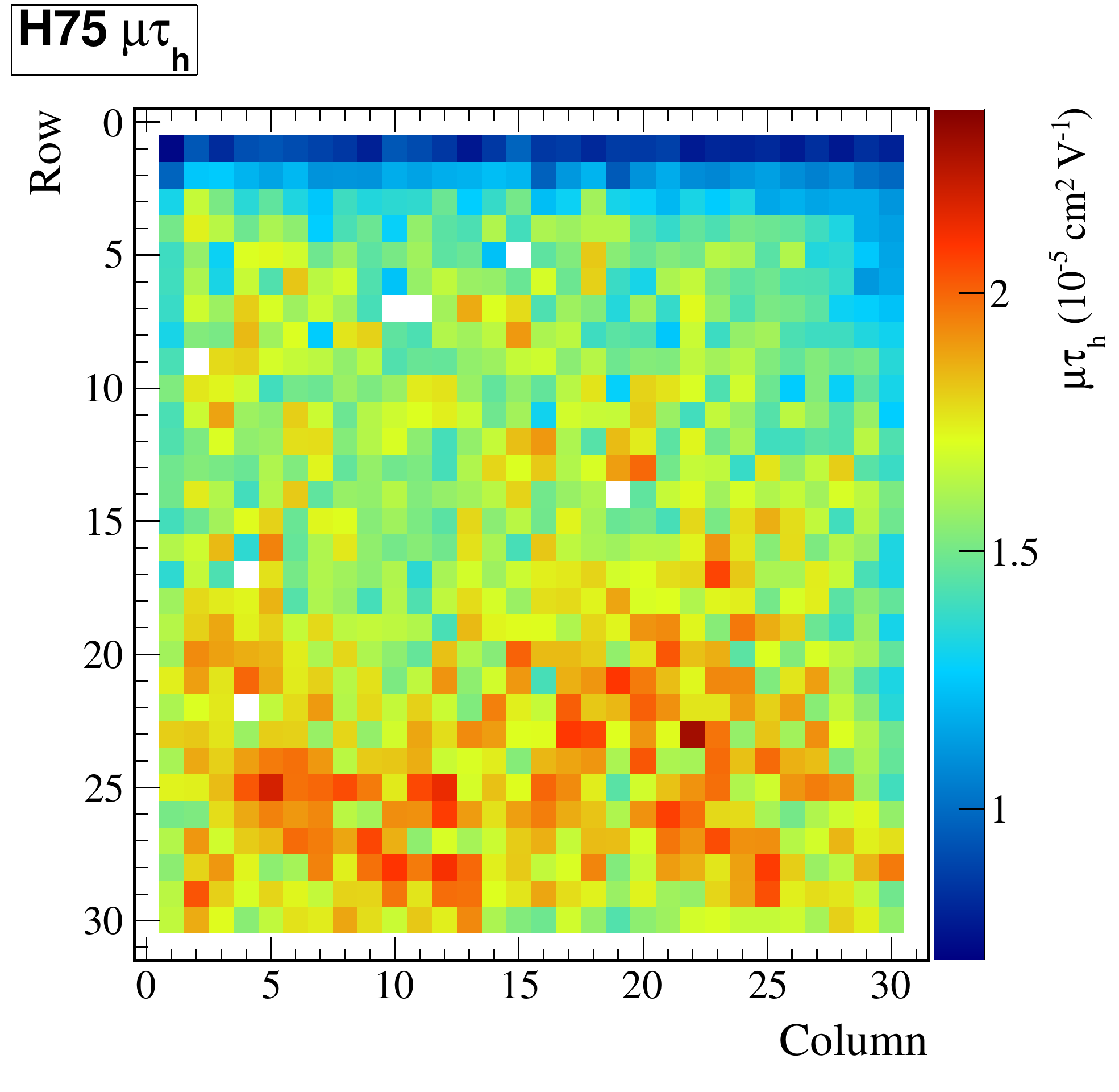}
   \end{tabular}
  \end{center}
  \caption{\label{fig:MapMuTauH}
  Same as figure~\ref{fig:MapMuTauE}, but obtained mobility-lifetime
  products for holes.}
 \end{minipage}
\end{figure}

\begin{figure}[ht]
 \begin{center}
  \begin{tabular}{c}
   \includegraphics[height=8.0cm]{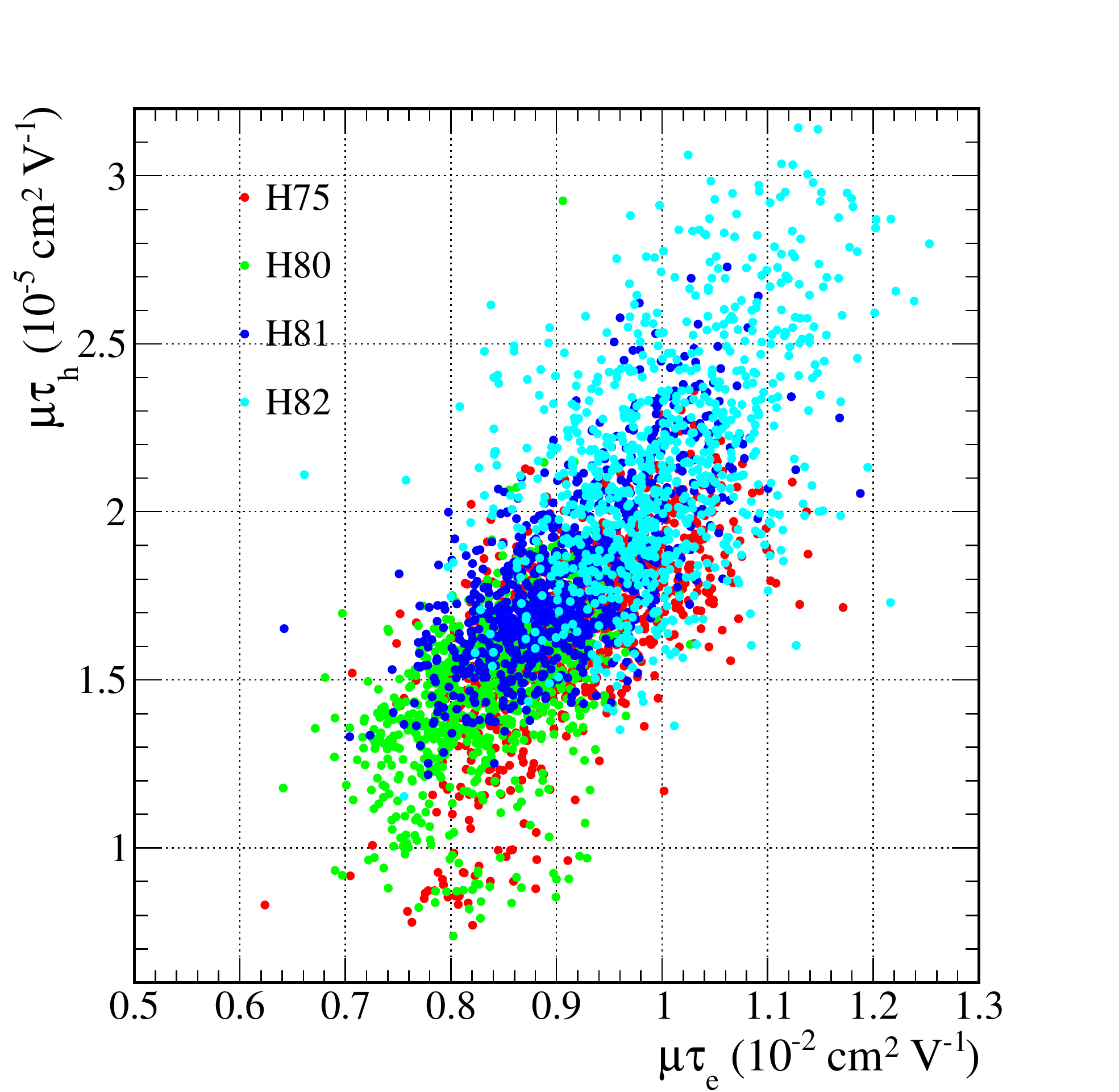}
  \end{tabular}
 \end{center}
 \caption{\label{fig:CorrMuTau}
 Correlation of measured mobility-lifetime products between electrons
 and holes.
 The red points represent the mobility-lifetime products for electrons in
 Figure~\ref{fig:MapMuTauE} and holes in Figure~\ref{fig:MapMuTauH}.}
\end{figure}

Figure~\ref{fig:MapMuTauE} and \ref{fig:MapMuTauH} show pixel maps
of mobility-lifetime products for electrons and holes,
respectively, obtained by analyzing the calibration data sets.
The mobility-life time product for electrons ranges
from $6.0 \times 10^{-3}$ to $1.2 \times 10^{-2}$~cm$^{2}$~V$^{-1}$,
while the mobility-lifetime product for holes ranges
from $7.0 \times 10^{-6}$ to $2.4 \times 10^{-5}$~cm$^{2}$~V$^{-1}$.
Typical errors are
3\% and 5\%
in the 1 $\sigma$ confidence level
for the mobility-lifetime products for electrons and holes, respectively.
The two $\mu\tau$ spatial maps are similar in pattern.
A correlation diagram of the mobility-lifetime products between
electrons and holes for 4 different CdZnTe hybrids is shown in
Figure~\ref{fig:CorrMuTau}.
Although these 4 crystals were prepared from different CdZnTe ingots,
the correction is modestly strong with a correlation coefficient of 0.72.
The strong correlation is reasonable because crystal defects suggest
mobility-lifetime products for electrons and holes should be related.

\section{Evaluation of Spectral Model}
\label{sec:eva}

In order to evaluate the accuracy of the NuSTAR detector model with the measured
charge transport properties, we obtained data during the instrument
thermal vacuum test, when the detectors were at the expected in-flight operating temperature,
by irradiating the calibrated CdZnTe hybrids
with the flight  $^{155}$Eu radioactive calibration source which is integrated into the
focal plane assembly.  The four hybrids were installed in the flight CsI shield.
By turning on the CsI shield, signals produced by charged particles and
$\gamma$-rays Compton-scattered with the shield can be eliminated using the
anti-coincidence scheme.   This event selection was incorporated into the model.

\begin{figure}[ht]
 \begin{center}
  \begin{tabular}{c}
   \includegraphics[width=10.5cm]{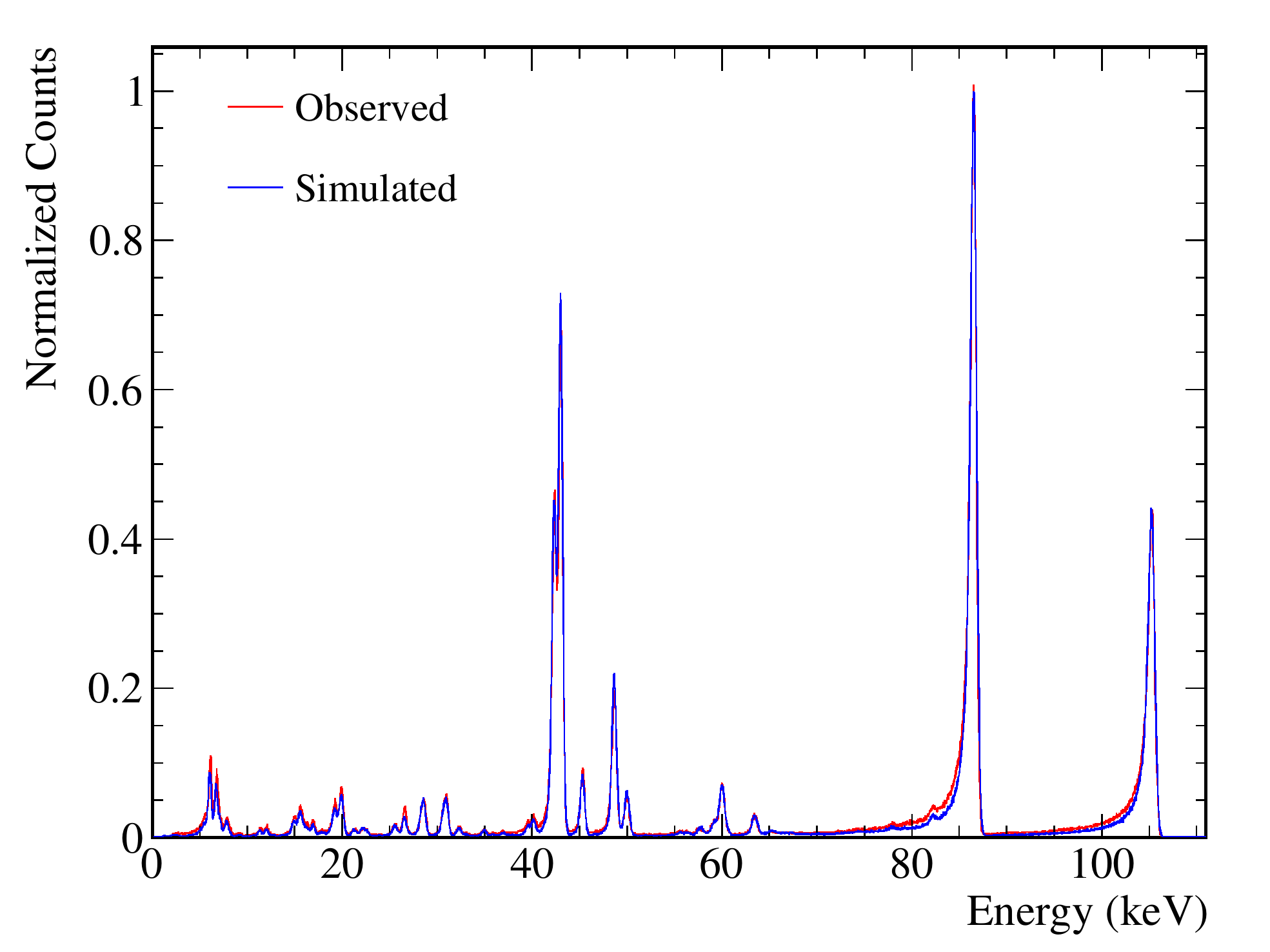}
  \end{tabular}
 \end{center}
 \caption{\label{fig:CompEu155}
 Comparison of the simulated $^{155}$Eu spectrum with the observed one.
 The detector is operated at 278~K temperature and -450~V high voltage.
 All of pixels are used  except for edge, corner, and dead pixels.
 Single-pixel events in which a trigger of the center pixel is solely
 activated are collected.
 The observed energy resolution in FWHM is 0.8~keV at 86.5~keV.
 Both spectra are scaled by the 86.5~keV peak counts.}
\end{figure}

We performed a
full Monte-Carlo simulation using the measured charge transport properties to generate a summed
spectrum model for the entire CdZnTe pixel array.
Figure~\ref{fig:CompEu155} compares the simulation result to the
real data.
The model can reproduce the
experimental data within about 10\% accuracy,  with the exception of the low-energy tail structures.  This is acceptable at
the pre-flight stage.
The model for two peaks around 29 and 31 keV, which are fluorescence
X-rays from the CsI shield excited by $\gamma$-rays from the $^{155}$Eu
source, is also in good agreement with the observed one.
This agreement demonstrates that the anti-coincidence model in the simulation
is accurate.
The disagreement of the tail components are possibly produced by an
extra charge loss near the gap surface between anode
pixels\cite{2002ITNS...49..270C}.
We need further studies to improve the model.

\section{Summary}

We have developed a  Monte-Carlo simulation code for the  NuSTAR CdZnTe
pixel detectors that generates a spectral model as a function of incident energy.
By comparing simulated models to observed data, we succeed in
extracting mobility-lifetime products for electrons and holes
independently for all $32 \times 32$ CdZnTe pixels.
We found a correlation (r = 0.72) between the
obtained mobility-lifetime products for electrons and holes as a function
of pixel.   The spectral response model we have developed for each individual
pixel can accurately reproduce calibration data taken on the flight detectors in the
laboratory.

\acknowledgments     

This work was supported under NASA Contract NNG08FD60C.
T.K. is supported by JSPS postdoctoral fellowships for research
abroad.  We thank Csaba Szeles for useful discussions.



  \end{document}